\documentclass[%
 aip,
rsi,%
 amsmath,amssymb,
 reprint,%
]{revtex4-1}

\usepackage{graphicx}
\usepackage{dcolumn}
\usepackage{bm,amsfonts}

\begin{document}

\preprint{AIP/123-QED}

\title[$mCIT$ for data-driven plasma science]{Microparticle cloud imaging and tracking for data-driven plasma science\footnote{Error!}}
\thanks{Invited talk presented to the 2nd International Conference on Data Driven Plasma Science (ICDDPS) 2019, Marseille, France, May 13-17, 2019.}

\author{Zhehui Wang}
\email{ zwang@lanl.gov (for correspondence).}
\affiliation{ 
Los Alamos National Laboratory, Los Alamos, NM 87545, USA
}%

\author{Jiayi Xu}%
\affiliation{%
The Ohio State University, Columbus, OH 43210, USA}%

\author{Yao E. Kovach}%
\affiliation{University of Michigan, Ann Arbor, MI 48109, USA}%

\author{Bradley T. Wolfe}
\affiliation{ 
Los Alamos National Laboratory, Los Alamos, NM 87545, USA}%

\author{Edward Thomas Jr.}
\affiliation{Auburn University, Auburn, AL 36849, USA}

\author{Hanqi Guo}
\affiliation{ 
Argonne National Laboratory, Lemont, IL 60439, USA}%

\author{John E. Foster}%
\affiliation{University of Michigan, Ann Arbor, MI 48109, USA}%

\author{Han-Wei Shen}%
\affiliation{%
The Ohio State University, Columbus, OH 43210, USA}%

\date{\today}

\begin{abstract}
Large data sets give rise to the `fourth paradigm' of scientific discovery and technology development, extending other approaches based on human intuition, fundamental laws of physics, statistics and intense computation. Both experimental and simulation data are growing explosively in plasma science and technology, motivating data-driven discoveries and inventions, which are currently in infancy. Here we describe recent progress in microparticle cloud imaging and tracking (mCIT, $\mu$CIT) for laboratory plasma experiments. Three types of microparticle clouds are described: from exploding wires,  in dusty plasmas and in atmospheric plasmas. The experimental data sets are obtained with one or more imaging cameras at a rate up to 100k frames per second (fps). Analyses of the time-dependent microparticle trajectories give time-dependent two-dimensional or three-dimensional information about the particle motion and ambient environment. The massive image and particle track data motivate development of machine-learning (ML) techniques for information extraction. A physics-constrained motion tracker, a Kohonen neural network (KNN) or self-organizing map (SOM), the feature tracking kit (FTK), and U-Net are described and compared with each other for particle tracking using the datasets. Particle density and signal-to-noise ratio have been identified as two important factors that affect the tracking accuracy. Fast Fourier transform (FFT) is used to reveal how U-Net, a deep convolutional neural network (CNN) developed for non-plasma applications, achieves the improvements for noisy scenes. The fitting parameters for  a simple polynomial track model are found to group into clusters that reveal the geometry information about the camera setup. The mCIT or  $\mu$CIT techniques, when enhanced with data models, can be used to study the microparticle- or Debye-length scale plasma physics. The datasets are also available for ML code development and comparisons of algorithms. %
\end{abstract}

\pacs{Valid PACS appear here}
\maketitle

\section{\label{sec:level1}Introduction}
Progress in experimental plasma physics, and similarly in almost any experimental and observational science, can be measured by the rate and the amount of new scientific data produced. How much data is produced in today's plasma physics experiments? Many plasma experiments, if not already, have the capability to produce 1 terabyte (TB) of data per day. The data rate is expected to grow due to the advances in instrumentation that parallel Moore's law for transistors. One scenario to obtain TB/day rate is through imaging of plasmas using video cameras. For each image file size of 1 megabyte (MB) and the video recording rate of 10$^2$ frames per second (fps), the data can be generated at 10$^8$ bytes/s. 1 TB of data can be collected in less than 3 hours per camera from a continuous plasma experiment. The state-of-the-art smart phone cameras, with an image size exceeding 10 MB per image and a constant streaming rate above 10  fps, can exceed such a data rate. In practice, this may not have been done often due to other considerations or limitations such as the bandwidth of the data transmission or the storage space available. Several commercial off-the-shelf high-speed cameras can deliver a data rate at 10$^4$ MB/s. In the exploding wire experiment shown below in Sec.~\ref{sec:PCIA}, 800$\times$1280 8-bit images were generated at 25 kfps. When such a high-speed camera is used in a pulsed 10-ms-long plasma experiment, an experimental duty cycle at 1 Hz can also produce 1 TB of image data within 3 hours. We shall limit our discussions to experiments and observations here. Computer simulations, especially large-scale simulations carried out on the supercomputers or large computer clusters, can readily exceed 10 TB/day data rate by orders of magnitude through parallelization. 

Another way to look at the large data sets is to estimate how much data is actually available from a plasma. The Bekenstein bound~\cite{Bek:1981} for the information content ($I_B$) from a plasma sphere with a radius $R$ is given by $I_B \leq 2\pi RE/\hbar c \ln2 = 2\pi c Rm/\hbar \ln2 \sim \alpha m R$ bits, with the coefficieint $\alpha = 2.577 \times 10^{43}$ bit/kg$\cdot$m.  Consider the ion and electron mass alone, for a 1-mm radius hydrogen plasma sphere with an electron density of 10$^{15}$ m$^{-3}$, the amount of data would be about 2.3$\times$10$^7$ TB. Since most of the internal mass energy does not change during a low temperature plasma measurement, we may replace the energy $mc^2$ by $3/2k_B(T_i+T_e) \sim$ 3 eV for three degrees of freedom for each pair of hydrogen ion and electron. The Bekenstein information content is reduced to 17 kB, which may be interpreted as the lower bound in the information content as the plasma system reaches the thermal equilibrium. When the system is deviated from the thermal equilibrium, more bits of information are needed to describe the system. Ignoring the internal degrees of freedom such as excited states of an ion, rotation and vibration states of a molecular ion, or nuclear spin states for now, only the position and momentum for each ion and electron are needed to fully describe the system. Similar to the `classical' or Shannon information content~\cite{Kri:2013}, an upper limit in information content is estimated to be on the order of $kN\log_2 N$ for $N$ electron and ion pairs. The multiplier $k=12$ is for 3 spatial dimensions, 3 momentum dimensions, and two types of particles (electrons and ions).

Collecting large data sets is not the ultimate goal of physical science including plasma physics, however. Until recently, experimental data have served multiple purposes: a.)~validation of theoretical or computational models; b.) characterization of an experimental or natural plasma; c.) validation or improvement of engineering designs or hardware. New applications of data for predictions are now emerging with large datasets. The new applications can come in different flavors such as 1.) Complex models with millions or more `adjustable parameters'; 2.) Adaptive models without explicit mathematical forms or parameterization; 3.) Real-time autonomous control of instruments and experiments. Such new applications, in particular, through statistical data models, can also complement or enhance established frameworks for predictions, whether they are based on the empirical laws or human intuitions, or the well-established theoretical framework of plasma physics such as MHD or kinetic theories, or extrapolation of computational results from gyro-kinetic simulations. In plasma physics, examples of difficult problems are quantitative interpretations of complex phenomena or observations such as solar corona mass ejection, or predictions such as the neutron yield or transients in a  fusion device, or engineering of new or better devices for semiconductor chip processing. The data-driven models based on TB and even larger data sets are called `machine-learning' models (MLM). On the one hand,  any MLM can be symbolically described by a functional relation {\bf Y}=$f$({\bf X}). Here {\bf X} stands for the inputs such as the raw images from a plasma experiment and {\bf Y} are the outputs such as the trajectories of particle motion or the growth rate of an instability. On the other hand, the derivation of unknown function $f$ involves heavily automated data pre-processing such as denoising and image transformations such as convolution, reduction (averaging and `pooling'), and extrapolations. If one represents each step by a `daughter' function, $f_i$, then $f(\cdot)=f_1(f_2(f_3 (... ( f_n(\cdot) )...)))$, where $n$ could be a large number that may not be able to be reduced to a simpler form, and therefore require automation or machine learning algorithms to generate. 

Neural networks (NN) have shown to be effective for processing large image data sets~\cite{GBC:2016}, solving differential equations, and simulation of physical systems~\cite{BNK:2019}. An NN is remarkable in recovering `empirical truth' when the physical framework for the data is unknown, and a generic approach with a large number of unknowns is used. In classification, the input data such as images need to be sorted into different categories (`dogs', `cats', `fish', `cars') as accurately as possible. Through parallel processing of different parts of an input image and passing the outputs in sync, it is  possible to process a  large-size image in real time quickly~\cite{CY:1988}. It has been shown that continuous functions can be approximated by feed-forward NN with a single hidden layer, which is the {\it universal approximation theorem}~\cite{Cyb:1989,HSW:1989}.  Eldan \& Shamir (2015) showed that, to approximate a specific function, a two-layer network requires an exponential number of neurons in the input dimension, while a three-layer network requires a polynomial number of neurons~\cite{ES:2015}. In using NN for stereo image analysis, for example, the state-of-the-art NN can recover millions of parameters out of a dataset. Learning polynomial functions with neural networks was described in~\cite{APVZ:2014}. Another generic example is to learn about probability density p({\bf X}) of a turbulent field given a certain number of samples (images). For a locally regular or continuous function p({\bf X}) and error $\epsilon$, an estimate $\tilde{p}({\bf X})$ that satisfies $\mathbb{E} (\| p - \tilde{p} \|) \leq \epsilon$, the number of samples required is given by~\cite{Mallat:online}
\begin{equation}
n \geq k \epsilon^{-d}.
\end{equation}
Since the dimension of a 1 meg-pixel image d$\sim $10$^6$, the huge number of samples $n$ required is also known as the curse of dimensionality. A school of thought is to find regularity properties (such as symmetry, scale separation) which can break the curse of dimensionality.


The datasets introduced here are related to microparticles or a microparticle cloud consisting of many microparticles. When interacting with plasmas, the theoretical framework developed for Langmuir probes are applicable when these particles can be treated as freely floating objects without a bias voltage. Dusty and atmospheric plasma data sets are included as examples. Laboratory dusty plasmas are amenable to video imaging diagnostics for several reasons. Individual micron-size dust particles are visible under laser illumination, which significantly enriches the observable phenomena compared with a plasma free of microparticles. A dusty plasma can be sustained in a steady state with relatively low RF power of a few watts. Consider a typical low temperature argon plasma with an electron temperature, $T_e$ = 3 eV and an electron density, $n_e =$  10$^{14}$ m$^{-3}$.  A 1-micron diameter silica particle in the plasma would acquire a negative charge of $q_d \sim$  2000 electrons.  Hundreds or more micron-size dust particles become suspended simultaneously in the plasma with an inter-particle separation about 100 times the particle size, or a dust density $n_d \sim 10^{10}$ m$^{-3}$. The time resolution of 10$^{-3}$ to 1 second is usually sufficient for many dynamic processes, determined by the characteristic time of 2$\pi/\omega_{pd}$ for collective dust motion, with $\omega_{pd} = \sqrt{q_d n_d/\epsilon m_d} \sim 20$ Hz, corresponding collective ion and electron would be in 100s of kHz and 10s of MHz regime respectively, requiring very high-speed cameras. Even when such a high-speed camera is available, it would still be difficult to see individual ions and electrons due to their smaller light scattering cross sections compared with the scattering from a micron-size particle. Imaging of microparticle interaction with plasmas can be extended to other plasmas including high-temperature magnetic fusion plasmas. When the core electron temperature can reach 10 keV and the edge temperature can be at least 10's of eV, there are additional complications of time-dependent microparticle size, shape and mass due to evaporation and sublimation. On the other hand, particle nucleation and agglomeration can lead to growth of micron and larger particles in an environment such as semiconductor and other material processing plasmas. In the simple scenarios when the particle size can be treated as a constant, outstanding physical questions related to the microparticle-plasma interactions include the microparticle charging, and the mechanisms behind the sophisticated motion patterns. 



The rest of the paper is divided into the following sections. Sec.~\ref{sec:frwk} gives an overview of microparticle tracking models motivated by physics and statistics, followed by Sec.~\ref{sec:dm1} on data models. Three tracking algorithms: Physics-constrained motion tracking, self-organizing map, and the feature tracking kit (FTK) are also explained in Sec.~\ref{sec:dm1}. Three types of microparticle clouds are described in Sec.~\ref{sec:dataset}: from exploding wires,  in dusty plasmas and in atmospheric plasmas. Sec.~\ref{sec:PCIA} compares results using different algorithms to process the image sets. Particle density and noise are found to be important factors that affect the effectiveness of the algorithms. An appendix is also included to reflect the fact that particle tracking and imaging are a rapidly expanding interdisciplinary field, and many tools are already available for beginners or new initiatives. 


\section{Microparticle tracking frameworks \label{sec:frwk}}
One application of microparticle cloud imaging and tracking (mCIT or $\mu$CIT) for plasma experiments is to infer physical properties such as particle shape, particle mass, electric charge, position, velocity, and/or force; {\it i.e.}, the `state or features of microparticles' in a plasma ambient. A collection of different particle trajectories can be further used to examine waves, instabilities, temperature, other kinetic and dynamic properties of the microparticles or the plasma ambient. The ultimate goal would be to fully deduce all the `features' that include both the microparticle's and plasma's properties such as plasma density and temperature profiles. Since the holistic framework with a comprehensive set of the plasma and microparticle properties as feature sets is beyond the scope of this work, we will only focus on a reduced set of features related to microparticle motion and dynamics for now. One purpose is to compare three types models for microparticle motion using the same experimental image data and furthermore, the availability of large data sets can potentially combine the three different approaches to extract more information about the microparticle-plasma system than by relying on each individual approach alone. The three approaches are physics-based models, statistical models and data-driven models. Sec.~\ref{sec:dm1} will elaborate on data-driven models. The second reason is that, advances in areas outside plasma research have led to a wealth of algorithms and computer codes. Adoption of these algorithms and computer codes for plasma applications can potentially be a fruitful area. Some examples will be given in Sec.~\ref{sec:PCIA}, following a summary of several microparticle datasets for imaging and tracking in Sec.~\ref{sec:dataset}.  

\subsection{\label{sec:pm} Physics models}
We start with physics-based particle tracking methods. Common use of digital cameras for particle imaging motivates discretized motion models, as governed by 
\begin{equation}
{\bf r}_{n+1} = {\bf r}_n + {\bf v}_n \Delta t.
\label{eq:rv}
\end{equation}
Here the subscripts $n+1$ and $n$ are for the consecutive time steps or video frame numbers with a time lapsed $\Delta t$ in-between them. 1/$\Delta t$ is the constant frame rate of the video camera. The bold face symbol {\bf r} is for the instantaneous position of a particle in three-dimensional (3D) physical space in general, which includes two-dimensional (2D) motion as a special case. {\bf v}$_n$ is the instantaneous velocity of the particle at the time step $n$, which in general also varies as a function of time or frame number $n$,
\begin{equation}
{\bf v}_{n+1} = {\bf v}_n + {\bf a}_n \Delta t.
\label{eq:steps}
\end{equation}

A physics-based tracking algorithm further prescribes ${\bf a}_n$ by Newton's equation for particle motion or equivalent through
\begin{equation}
M_n {\bf a}_n= \sum_i {\bf f}^i_n.
\label{eq:newton}
\end{equation}
Here $M_n$ is the particle mass at the time step $n$ and the summation on the RHS is for different forces. In a laboratory plasma, the sum may include gravity, neutral gas drag, ion drag force, electrostatic or electromagnetic force for a charged particle, {\it etc}. In the strongly coupled regime, electrostatic Coulomb interactions among neighboring particles also have to be included~\cite{ORDS:2012}. In cases when the particle mass $M_n$ varies with $n$, the RHS can also include a `rocket force' given by {\bf v}$_n$ d$M_{n}$/dt = {\bf v}$_n$ ($M_{n+1} -M_n$)/$\Delta t$. We mention without further elaboration that Eqs.~(\ref{eq:rv}), (\ref{eq:steps}) and (\ref{eq:newton}) can also be extended to include additional degrees of freedom such as particle rotation or spin since a particle consists of millions or more atoms. 

To learn about the plasma conditions or unknown physical properties such as the electric charge or a force on a particle through Eq.~(\ref{eq:newton}), it will require the information of {\bf v}$_n$ and {\bf a}$_n$ first through Eqs.~(\ref{eq:rv}) and (\ref{eq:steps}). Therefore, it appears that a non-physical approach such as a data-driven method would be a prerequisite to derive {\bf r}$_n$ and {\bf r}$_{n+1}$. Meanwhile, even an initial estimate of particle velocities based on physics arguments such as particle kinetic energy, momentum conservation, or energy conservation would be useful to correctly pair up ${\bf r}_n$ with ${\bf r}_{n+1}$. One example will be included in Sec.~\ref{sec:dm1} and Fig.~\ref{fig1:motionT20} below. Another reason why a pure data-based method may not be the best option is related to the so-called NP-hard problems in computing. When an image contains many microparticles, correctly pairing of the microparticles from one image to another is not obvious, giving rise to the `particle linking' problem in tracking algorithms. For example, when processing a video with 100 particles per frame and 100 frames long, a random pairing algorithm would give rise to $\sim$100$^{100}$=10$^{200}$ possible tracks. Data association and track-to-track association, two fundamental problems in multi-target tracking, are instances of an NP-hard combinatorial optimization problem known as the multidimensional
assignment problem (MDAP)~\cite{EPER:2018}. Physics constraints will allow substantial reduction of computing time to correctly identify $\sim 100$ tracks. Physical considerations also motivate more `feature learning' such as the particle size, particle brightness from raw data for effective algorithm development. Another application of physics models would be to provide `ground truths' for data model training.

In raw image data, {\bf r}$_n$ and {\bf r}$_{n+1}$ are not measured directly. An optical camera image is a 2D projection of a 3D physical scene. Such a projection is usually described by epigeometry~\cite{Zhang:1998,HZ:2004,Wang:2016}, which relates 3D coordinates {\bf r} (suppressing the subscript for now) in the physical space with at least two independent 2D projections or two pairs of camera coordinates {\bf q}$_+$ ($u_+$, $v_+$) and {\bf q}$_-$ ($u_-$, $v_-$). A triangulation algorithm can recover {\bf r} from {\bf q}$_\pm$. Therefore, in addition to linking of particles from different image frames in a single camera, another type of linking algorithm is required to link particles from a pair of cameras which project the same physical scene from different positions and angles. 

Derivation of the positions {\bf q} or ($u$, $v$) from an image is called {\it particle localization}. Subscripts $\pm$ are suppressed to avoid clutter.  A particle image is typically spread over a cluster of neighboring pixels with an intensity distribution $I_j(\tilde{{\bf q}}_j)$. Isolation of the pixel cluster from the rest of the image is called {\it instance segmentation}. Once segmented, one common algorithm to locate {\bf q} is through the centroid of the intensity cluster,  {\bf q} = $\sum_j I_j \tilde{{\bf q}}_j /\sum_i I_i$. Due to the finite pixel size of a camera, signal-to-noise ratio of the particle image intensity, motion blur due to the finite camera exposure time, and particle image overlap in high particle densities, {\bf q} can only be determined within a certain accuracy. The errors of {\bf q} measurement can propagate down through the whole data processing chain that essentially limit the accuracy about a derived physical quantity such as the spatial coordinates and velocities. 


\subsection{\label{sec:sm} Statistical models}

In either a data-based approach or physics-driven approach, a central problem is about effectively dealing with noise or uncertainties~\cite{FGL:2011}, which naturally motivates statistical models for positions {\bf r}$_n$ and other physical quantities as defined in Sec.~\ref{sec:pm}. A statistical tracking model predicts the motion of an object such as its next position {\bf r}$_{n+1}$ as a probabilistic distribution function, P({\bf x}$_{n+1} |$ {\bf q}$_1$, {\bf q}$_2$, $\cdots$, {\bf q}$_n$). Here ${\bf q}_i$ with $i=1, \cdots, n$ are the measured quantities as a function of time up to the $n$-th step. One example of ${\bf q}_i$ is the image coordinates {\bf q}$_\pm$ as given above. Each ${\bf q}_i$ can be further framed as a function of ${\bf x}_i$ as explained in Eq.~(\ref{eq:m1}) below. ${\bf x}_{n+1}$ is a generalized state vector from the three-dimensional position {\bf r}$_{n+1}$, that may also include for example instantaneous velocity, acceleration, electric charge ($Q$); {\it i.e.}, {\bf x}$_{n+1}$ =({\bf r}$_{n+1}$, {\bf v}$_{n+1}$, {\bf a}$_{n+1}$, $Q_{n+1}$, $\cdots$).  Based on the probabilistic distribution function, and the new measurement {\bf q}$_{n+1}$, the optimal estimate for {\bf x}$_{n+1}$, or $\hat{{\bf x}}_{n+1}$ can be made. The new information through {\bf q}$_{n+1}$ also updates the probabilistic function to P({\bf x}$_{n+2} |$ {\bf q}$_1$, {\bf q}$_2$, $\cdots$, {\bf q}$_n$, {\bf q}$_{n+1}$), which allows optimal prediction of the future state, $\hat{{\bf x}}_{n+2}$ with additional measurements {\bf q}$_{n+2}$ and so on. 

One well-known statistical model with applications to particle tracking is the Kalman filter~\cite{Kal:1960,Jaz:1970,AM:1979,MC:2011}. Originated in the sixties for object tracking using radar, sonar and electromagnetic techniques, recent surges in automated object recognition and tracking are motivated by computer vision and image processing applications such as robotics and self-driving cars~\cite{HMH:2006}. Kalman filter is an optimal recursive Bayesian filter for linear functions subjected to white Gaussian noise. Constant gain Kalman filter is computationally fast for single-object detection, tracking, and localization. The original Kalman filter has been extended in various ways. The extended Kalman filter (EKF) and unscented Kalman filter are nonlinear versions of the Kalman filter~\cite{WvdM:2000,BLK:2001,Ral:2010}. Additional nonlinear, non-Gaussian filters and application examples can be found in the books~\cite{RAG:2003,MC:2011} and citations therein. Example applications of Kalman filters and variants related to microparticles in plasmas can be found in Ref.~\cite{HMH:2006,ORDS:2012,FGH:2016}.

A Kalman-filter statistical model modifies Eqs.~(\ref{eq:rv}), (\ref{eq:steps}) and (\ref{eq:newton}) by a state equation~\cite{FP:2011},
\begin{equation}
{\bf x}_{n+1} = {\bf A}_{n} {\bf x}_n + {\bf f}_n+ {{\boldsymbol{\epsilon}}_{\bf x}}_n,
\end{equation}
and a measurement equation
\begin{equation}
{\bf q}_n = {\bf B_n}{\bf x}_n + {{\boldsymbol{\epsilon}}_{\bf q}}_n
\label{eq:m1}
\end{equation}
at the time step $n$. The state equation describes the state evolution to the next step. The measurement equation relates the state ${\bf x}_n$ to the measurement ${\bf q}_n$. The term ${\bf f}_n$ stands for external control including an external force.  ${{\boldsymbol{\epsilon}}_{\bf x}}_n$ and ${{\boldsymbol{\epsilon}}_{\bf q}}_n$ symbolize uncertainties or noise in the state evolution and errors in measurements respectively. Both ${{\boldsymbol{\epsilon}}_{\bf x}}_n$ and ${{\boldsymbol{\epsilon}}_{\bf q}}_n$ have a zero mean and non-vanishing covariance. A common noise and error model is the white Gaussian noise model~\cite{Jaz:1970,AM:1979,MC:2011}, when ${{\boldsymbol{\epsilon}}_{\bf x}}_n$ and ${{\boldsymbol{\epsilon}}_{\bf q}}_n$ have a Gaussian distribution with a zero mean.

The Kalman filter recursive algorithm can be summarized as
\begin{equation}
\hat{{\bf x}}_{n+1} = {\bf A}_{n} {\bf x}_n + {\bf f}_n+ {\bf K}_n ({\bf q}_n - {\bf B_n}{\bf x}_n),
\end{equation}
\begin{equation}
{\bf K}_n = P({\bf x}_n ) {\bf B}_n^T [{\bf B}_n P ({\bf x}_n) {\bf B}_n^{T} + E({{\boldsymbol{\epsilon}}_{\bf q}}_n) {\bf I}]^{-1},
\end{equation}
\begin{eqnarray}
P({\bf x}_{n+1}) &=& {\bf A}_{n} P({\bf x}_n) {\bf A}_{n}^T + E({{\boldsymbol{\epsilon}}_{\bf x}}_n) {\bf I} \\
&&  - {\bf A}_n K_n [{\bf B}_n P({\bf x}_n) {\bf B}_n^T +E({{\boldsymbol{\epsilon}}_{\bf q}}_n) {\bf I} ] K_n^T {\bf A}_n\nonumber
\end{eqnarray}
which are supplemented by the initial conditions.




The extension of the state variable into including multiple particles is sometime desirable, for example, when the mutual interactions among different particles exists. The probability hypothesis density filter is an extension of the single-target Bayesian framework to multiple targets. The method recursively estimates a multi-target state from observations, propagating the first-order moment of the multi-target posterior~\cite{Mah:2003}. Multiple particle tracking is now also used for extended object tracking. Another Kalman filter variant for tracking multiple objects is particle filters~\cite{HLP:2001,GGB:2002,WLSC:2017}.

\section{Data models \label{sec:dm1}}
Both physics models and statistical models rely on assumptions to make predictions. Good assumptions play multiple roles such as {\it a}. simplifying the calculations or the reasoning process; {\it b}. supplementing the incomplete knowledge about the physical systems; and {\it c}. fitting the experimental data well. In practice, good and simplifying assumptions can be difficult to come by in complex plasmas and as a result, the predictive power of a physics or statistical model is limited. For example, in most physics or statistical frameworks for microparticle interaction with plasmas, microparticles are almost always assumed to be a perfect sphere, which reduces the number of geometrical parameters to one (particle radius). In another example, models for material properties rarely consider the surface morphology that could significantly affect the electron emissivity, electron scattering and trapping, and therefore the amount of electric charge on a microparticle immersed in a plasma. Kinetic effects, non-equilibrium states of a plasma and especially near the sheath of a microparticle is another example when good physics and statistical model can be difficult to develop. Simplifying assumptions are sometimes essential to avoid computational penalty associated with sophisticated physics or statistical models. The continuous decrease in computing cost, which has given rise to data models, certainly opens door to more sophisticated physics and statistical models.

A data model, such as the least squares fitting or regression, complements physics models and statistical models without relying on the assumptions discussed above. Since humans are essential behind the assumptions, doing away with the assumptions paves the way towards full automation without human intervention~\cite{ZYY:2013}. This is still in the early phase of development. In spite of their different designs, the common measure of efficacy for either a data model, or a physics model, or a statistical model, or their hybrids, is to validate its predictive power with new experimental data. 

A data model may be characterized by its parameters. A simple data model such as a linear regression only has two parameters. A modern deep convolutional neural network may have millions or more parameters. The data that are used to find out or tune the parameters are called `training data'. New data can be used to validate or further tune the model parameters.  Independent of the model complexity, the parameters may start with random initial values. Least squares fitting is a widely used procedure to tune the model with two or a few parameters. Backpropagation is a procedure developed for parameter tuning in neural networks.


We examine data models that determine the individual particle coordinates, velocities and acceleration, or particle tracking velocimetry (PTV). A closely related class of approaches, particle image velocimetry (PIV), generally does not need to determine the particle coordinates explicitly and is usually used when the particle density is high and separation of individual particles is difficult.  The algorithm needs to perform two basic functions: 1. Particle localization; 2. Particle matching or linking. Particle localization is to determine the coordinates of a particle. Particle matching or linking is to recognize the same particle at different times or from different camera views. A `particle' can be a macroscopic object such as a star~\cite{SM:2009,KB:2016}, a car or a human, or a microscopic object such as a biological cell or an organelle inside the cell. Despite of different physical origins, the images captured have similar features through the uses of telescope, microscope and other hardware. The similarities in the images potentially allow data models developed for one type of objects for another.  

Neural network methods for particle tracking (localization and linking) have been investigated by a growing number of authors~\cite{GP:1995, Lab:2000}. Different types of artificial neural networks has been used for tracking: feedforward neural networks (such as autoencoder, convolutional)~\cite{HTS:2016}, feedback neural networks, recurrent neural networks (such as Hopfield, long short-term memory, competitive or self-organizing), radial basis function neural network,  modular neural network. Additional examples are given in TABLE~\ref{tab:table1}.

\begin{table}[thb]
\caption{\label{tab:table1} Examples of data-driven tracking algorithms and its applications. Illumination is assumed to be optical by default and otherwise specified. {\it CC}: Cascade classifier.  {\it CNN}: convolutional neural network. {\it DM}: diffusion maps. {\it HF}: Haar features. {\it KNN}: Kohonen neural network. {\it LSTM}: Long Short Term Memory. {\it RNN}: recurrent neural network. {\it SOM}: self-organizing map.}
\begin{ruledtabular}
\begin{tabular}{lccc}
{\bf Instrument} & {\bf Application} &{\bf Parameter} &{\bf Data Model}\\
(Image set) & &(feature)& \\
\hline
(simulations)~\cite{Lab:2000} & fluids & position & {\it KNN}\\
& &  & ({\it SOM})\\
microscope~\cite{JZKW:2007} & cellular & position & {\it HF} \\
&dynamics& & \\
microscope~\cite{LZGF:2015} & self-assembly & cluster & {\it DM} \\
&&(distance)& \\
video~\cite{SA:2015} & surveillance  & object & {\it CNN}+{\it RNN} \\
cryo-EM~\cite{ZOM:2017} & macromolecules  & object & {\it deep CNN} \\
particle & high-energy & track & {\it LSTM}+{\it CNN} \\
detector~\cite{TAB:2018}&physics&&\\
holograms~\cite{HAOG:2018} & colloidal &  3D position & {\it CC},{\it CNN}\\
 & science &  &  \\
 MNIST~\cite{LCL:2018} & computer &  position & {\it SO-Net}\\
 & science & cloud &  \\\end{tabular}
\end{ruledtabular}
\end{table}

In particle tracking, the input vectors {\bf I} are the intensity maps or images. The output vectors {\bf y} can vary. For particle localization, {\bf y} are the coordinates of individual particles. Correspondingly, NN algorithm workflow could be as simple as {\bf I} $\rightarrow$ {\bf q} = {\bf y}. Here the output {\bf y} is the raw camera coordinates {\bf q}, using the symbols defined in Sec.~\ref{sec:frwk}. Additional steps or NN layers can be added, for example, {\bf I} $\rightarrow$  {\bf I}$'$ $\rightarrow$ {\bf q} $\rightarrow$ {\bf r} = {\bf y}. Here {\bf I}$'$ stands for tranformed raw images after denoising, smoothing, convolution, {\it etc}. In a recent example~\cite{NSL:2018}, output is the probability of a particle centered at a pixel. For particle linking, {\bf y} correspondsn to linkage between the coordinates from different images. For example, if two pairs of coordinates {\bf q}$_1$ and {\bf q}$_2$ are linked, then {\bf y}({\bf q}$_1$, {\bf q}$_2$) = 1; otherwise {\bf y}({\bf q}$_1$, {\bf q}$_2$) = 0. 

In general, the input and output can be described by an unknown function $f$: {\bf I} $\rightarrow$ {\bf y}, and modeled by a neural network for a set of given data pairs $\{ ({\bf I}, {\bf y}) \}$.  The building elements of a neural network are neurons which are connected to other neurons~\cite{Bishop:2006}. Each neuron calculates a weighted sum of the outputs of neurons which are fed to it (and usually adds a bias term, $b$),
$z = \theta \left(\sum_i w_i x_i + b \right)$,
here $\theta$ is called the activation function.  $x_i$ and $z$ are the individual neuron inputs and output. Some of the popular activation functions include rectified linear unit (ReLU), tanH, softmax. In a feed-forward neural network, the neurons are arranged in layers, and the outputs of each layer forms the inputs to the next layer.  Each layer may be interpreted as a transformation of its inputs to outputs. As the input data propagates through the layers, higher-level concepts or features emerges. The depth of a network is the number of layers and the size of the network is the total number of neurons. In a convolutional neural network (CNN), the weighted input sum to a neuron through multiplication $\sum_i w_i x_i = {\bf w} \cdot {\bf x}$ is replaced by a convolution operation $\otimes$: ${\bf w} \otimes {\bf x}$. The weight parameter {\bf w} in a CNN is also called a kernel. Further discussions about algorithms are given in the Appendix~\ref{appen:1}.

Once the neuron linkage and activation functions within NN are chosen by design, the next step is to determine the values of the weights $\{ {\bf w} \}$ and biases $\{ b \}$ through so-called learning process using training dataset. Training is iterative starting with random values (within certain bounds) for $\{ {\bf w} \}$ and $\{ b \}$. A cost function such as $L_2$ norm $\|y_i-f_i\|_2 \equiv (y_i -f_i)^2$ can be obtained~\cite{Bishop:2006}, where $y_i$ is the expected output and $f_i$ is the corresponding value from NN. Iteration continues by adjusting $\{ {\bf w} \}$ and biases $\{ b \}$ to minimize the cost function until the desired value or error is reached. Gradient descent and backpropagation algorithms have been developed for the iteration~\cite{Wer:1994,RHW:1986a, RHW:1986b}. It has been shown that sophisticated NN such as deep CNN involving millions or more parameters $\{ {\bf w} \}$ and $\{ b \}$ are computationally hard to train~\cite{LSS:2014}. In  practice,  recent NN  are  commonly trained  using stochastic gradient descent (SGD) for expediency and a variety of tools are used that include proper selection of activation functions ({\it e.g.} ReLU), over-specification ({\it i.e.}, train networks which are larger than needed), and regularization~\cite{Bishop:2006,GBC:2016,ANg:2019}.

\subsection{Physics-constrained motion tracking  \label{pcmt}}
\begin{figure}[hbt]
\includegraphics[width=2.5in, angle=0]{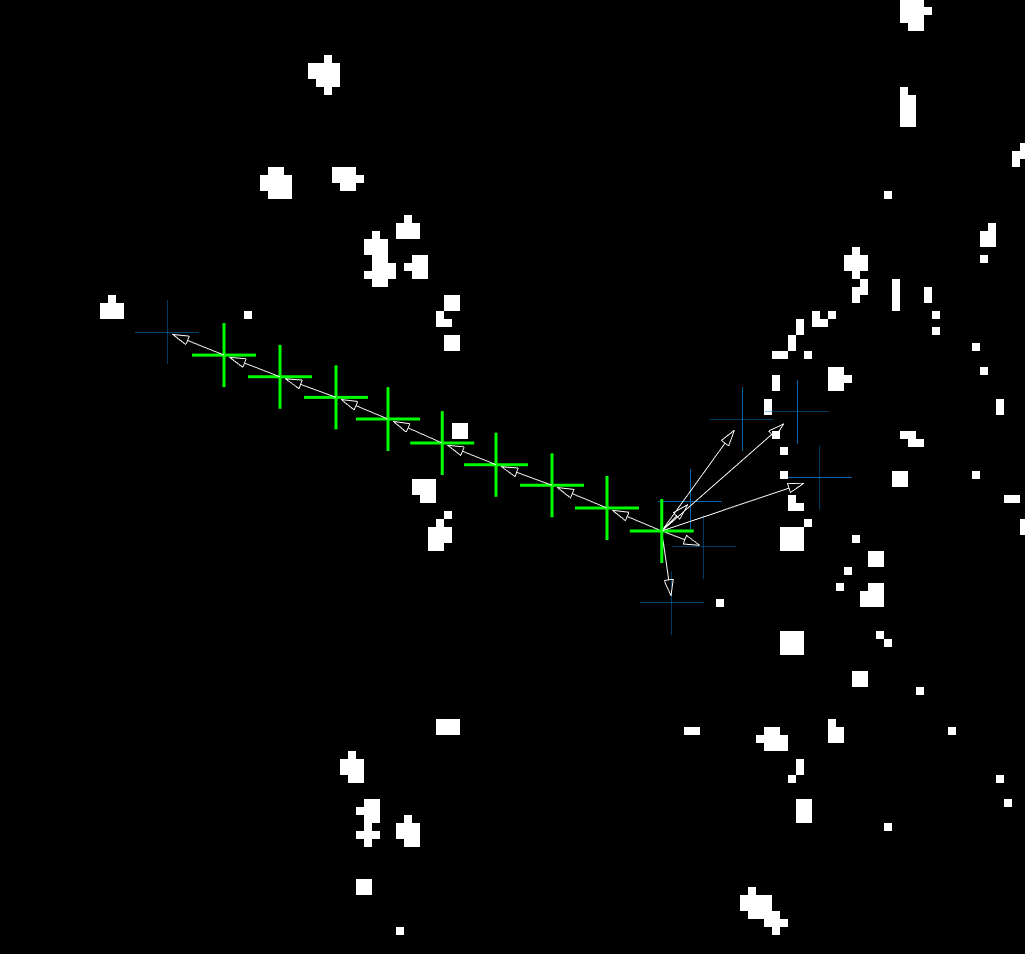}
\caption{ illustration of a motion tracking algorithm. The image background corresponds to particles in the last step. Initial search radius = 20 pixels. error in search = 2 (determined by particle density).}
\label{fig1:motionT20}
\end{figure}

We first describe a motion-based linking algorithm using the local constant velocity approximation. Following particle localization, the sequence of camera coordinates for individual particles is created from each camera. One type of linking algorithm is needed to assign the particle coordinates to different tracks in the camera plane. For stationary particles that do not move, the track reduces to a point. The essential function of the algorithm is to link particles from two different images, corresponding to particle positions at two different times. For a given particle in one image, estimate of the particle velocity and the time lapse between the two images give the estimate of the search radius ($R_s$). Since the direction of the motion is unknown, the candidate particles within the search radius now give the possible velocity vectors (both magnitude and direction), as illustrated in Fig.~\ref{fig1:motionT20}. A third and additional images can be used to down-select the particles. In Fig.~\ref{fig1:motionT20}, 7 candidate particles are found in the initial search. A third image is sufficient to settle down on the correct search, and further confirmed by sequence of 6 additional frames. 

\begin{figure}[hbt]
\includegraphics[width=2.5in, angle=0]{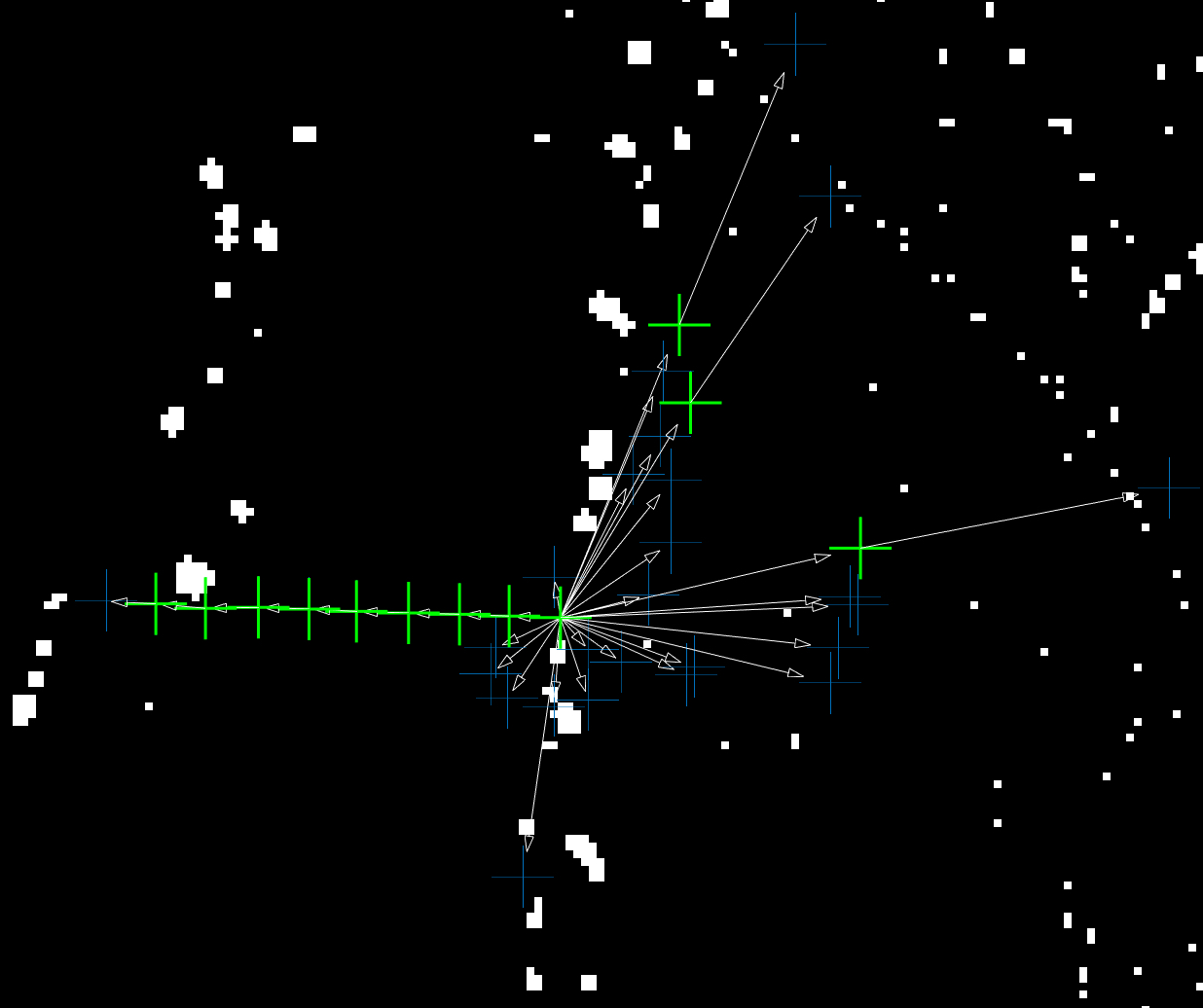}
\caption{ illustration of the motion tracking algorithm with twice the initial search radius as in Fig.~\ref{fig1:motionT20}.}
\label{fig1:motionT40}
\end{figure}

Only a rough estimate of the search radius $R_s$ is needed as illustrated by Fig.~\ref{fig1:motionT40}, when the search radius is doubled from Fig.~\ref{fig1:motionT20}.  There are now 25 possible matches within $R_s$. A subsequent third image reduces the candidates to 4. A 4th image is sufficient for the final answer.  In both examples shown here in Fig.~\ref{fig1:motionT20} and Fig.~\ref{fig1:motionT40}, a nearest-neighbor algorithm would give the incorrect linking. The nearest-neighbor approach fails here because of large particle density ($n_p$), complicated further by the fact that the velocities of the particle motion ($v_p$) are sufficient large so that 
\begin{equation}
l_p \equiv v_p \Delta t > n_p^{-1/3},
\end{equation}
where $\Delta t$ is the time lapse between the images. A nearest-neighbor algorithm would work if $l_p < n_p^{-1/3}$.

\subsection{KNN/SOM tracking \label{knnsom}}
Similar to Ref.~\cite{GP:1995, Lab:2000}, we describe a parallel tracking method using Kohonen neural network (KNN), also known as the self-organizing map (SOM)~\cite{Koh:1998, Koh:2001}.  
Here the KNN consists of three layers: the input layer, the neuron layer, and the output layer. The $N$ inputs are the individual particle 2D positions, $\{{\bf q}_1, {\bf q}_2, \cdots, {\bf q}_N \}$ from one of the two frames. The output layer is the possible matched particles from the other frame, $\{{\bf q'}_1, {\bf q'}_2, \cdots, {\bf q'}_M \}$. $M \neq N$ in general. For a particle that occupies multiple pixels, the average position or the centroid that averages the pixel coordinates is used, similar to Sec.~\ref{pcmt} above.

\begin{figure}
\includegraphics[width=2.0in, angle=0]{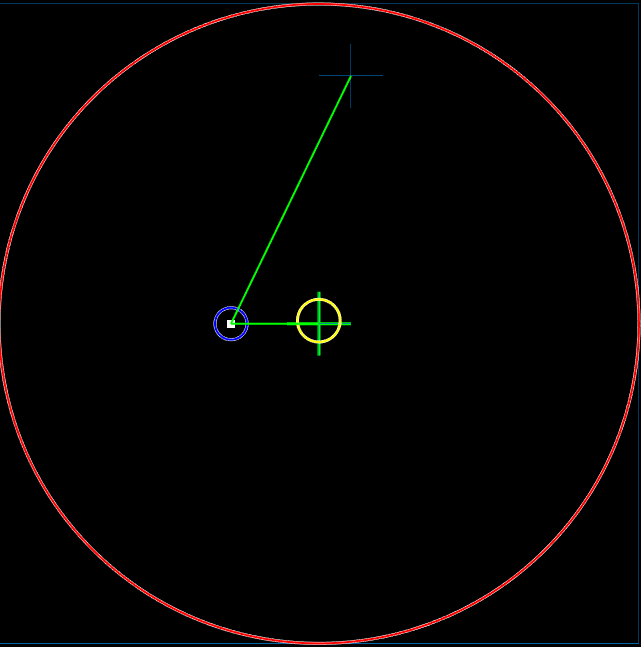}
\caption{The KNN/SOM algorithm reduces to the nearest neighbor algorithm when the particle density is low. In this example, only one neighbor exists within a predefined initial search radius $R_s^{(1)}$.}
\label{fig1:KNN}
\end{figure}

Each neuron (${\bf z}_i$, $i$=1, 2, $\cdots$, $N$) assumes one of the $N$ initiation positions $\{{\bf z}_i^{(1)} ={\bf q}_i$ , $i =1, 2, \cdots, N \}$ to start. The subsequent positions at the $(k+1)$th step evolve from the $k$th step through  {\bf z}$_i^{(k+1)}$ = {\bf z}$_i^{(k)}$ + $\sum_{j=1}^M {\bf w}_{ij}^{(k)}$ with the weight ${\bf w}_{ij}^{(k)}$ given by~\cite{Lab:2000} 
\begin{equation}
{\bf w}_{ij}^{(k)} = \alpha ({\bf q'}_{j} - {\bf z}_{ci}^{(k)}) H(R_s^{(k)} - |{\bf z}_{i}^{(k)} - {\bf z}_{ci}^{(k)}| ),
\label{eq:Kal}
\end{equation}
where the summation is over the possible matches $M$ in the other image frame. $0 < \alpha <1$ is a constant and we use $\alpha = 0.1$ here. ${\bf z}_{ci}^{(k)}$ is the `winning' neuron position that is the closest to {\bf q}$'_{j}$ at the step $k$. $H (x)$ is the Heaviside step function that satisfies $H(x) =1$ for $ x > 0 $ and $H (x) =0$ otherwise. $R_s^{(k)}$ is the search radius at the $k$th step. In the low density particle regime, KNN reduces to the nearest neighbor algorithm, as shown in Fig.~\ref{fig1:KNN}.

\begin{figure}
\includegraphics[width=2.0in, angle=0]{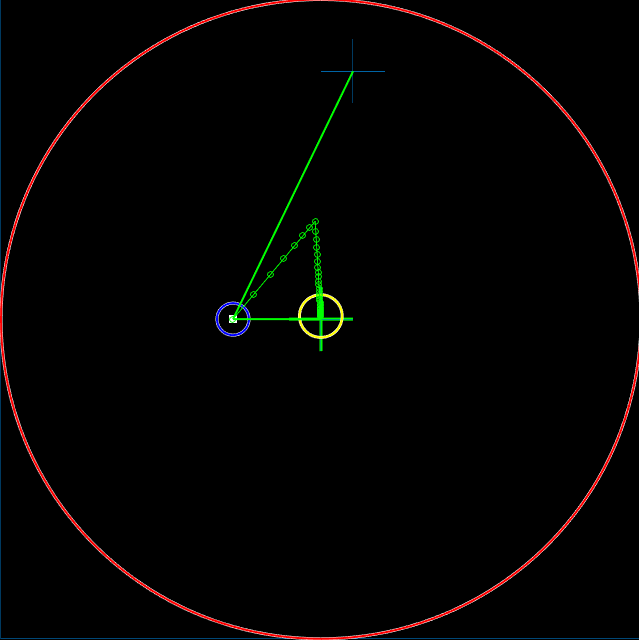}
\caption{ Illustration of the KNN algorithm described by Eq.~(\ref{eq:Kal}) when there are at least two particles falling within the initially search radius $R_s^{(1)}$. The neuron moves towards the weighted centroid between the two point, until the search radius $R_s^{(k)}$ shrinks and only one particle is nearby. Then the neuron is `attracted' to the object, similar to the nearest neighbor algorithm.}
\label{fig2:KNNAlgo}
\end{figure}

When the particle density increases, KNN is different from the nearest neighbor algorithm as illustrated in Fig.~\ref{fig2:KNNAlgo}, where we select a case when two neighbors in the other frame are found within the initial search radius $R_s$. The neuron starts to evolve in-between the two possible matches until the search radius $R_s$ shrinks to the point when only one nearest neighbor is found. Then it evolves along the straight line determined by the neuron position and the matched particle position. KNN/SOM differs from the nearest neighbor algorithm and the physics-based tracker by design. It is attractive to use KNN/SOM for parallel tracking of multiple particles. Meanwhile, its effectiveness varies significantly with the particle density and further studies will be needed to explain the effectiveness of the algorithm.

\subsection{The feature tracking kit (FTK) \label{sec:ftk}}
The feature tracking kit (FTK)\cite{XuG:2019} is a general purpose library to track features in both simulation and experimental data in a scalable manner.  The motivation of FTK is to ease the burden of developing domain-specific algorithms to track features such as local extrema and superlevel sets.  Basically, FTK incorporates an ensemble of techniques including machine learning, statistical, and topological feature tracking algorithms to help scientists define, localize, and associate features over space and time.  FTK has a unique design to generalize existing 2D/3D feature descriptors to trace the trajectories of features in 3D/4D spacetime directly.  FTK also supports feature tracking in distributed and parallel machines.

Specific to the microparticle tracking application, we use FTK to track local maxima in the time-varying imaging data.  We first derive the image gradients, and then localize maxima -- locations where the gradient vanishes and the Jacobian is negative definite in the 3D spacetime mesh.  Based on our scalable union-find data structure~\cite{jiayi:2019}, we then associate these space-time maxima and construct their trajectories based on the connectivities of spacetime mesh elements. Examples of trajectories obtained from FTK are given below in Fig.~\ref{fig:ftk1}.

\section{Experimental image datasets \label{sec:dataset}}
Here we summarize three experimental video datasets of microparticle clouds: in an exploding wire experiment, in a dusty plasma and in an atmospheric plasma. The exploding wire dataset has the highest signal-to-noise ratio among the three. The dusty and atmospheric plasma datasets provide examples of rich motion patterns at individual particle level as well as particle group level when particles are immersed in plasmas.

\begin{figure}[htb]
\includegraphics[width=2.5in, angle=0]{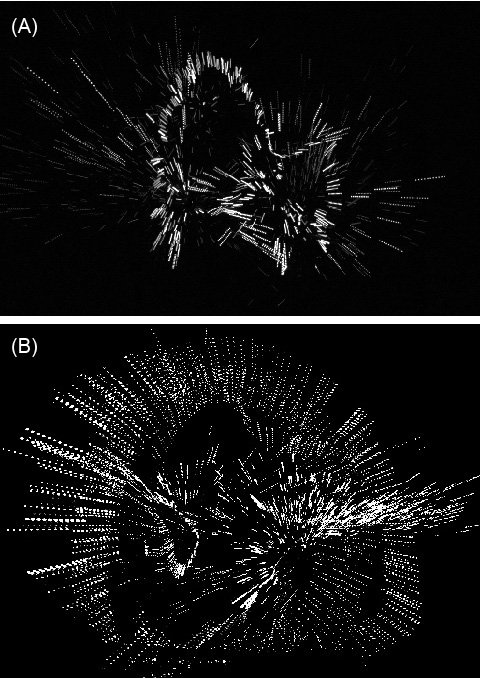}
\caption{ Two examples of microparticle pattern created from exploding wires. Each panel is a superposition of 11 consecutive video frames taken at the rate of 25k fps. The top panel (A) has a lower particle density than the one below.}
\label{fig:EW1}
\end{figure}

The exploding wire experimental setup and some analyses have been reported in~\cite{Wang:2016} and subsequent publications~\cite{CWW:2018,SWR:2018,WIC:2018}. Two new examples of microparticle clouds are shown in Fig.~\ref{fig:EW1}  at different particle densities. 

\subsection{Dusty plasmas in a strong magnetic field}
\begin{figure}[htb]
\includegraphics[width=3.5in, angle=0]{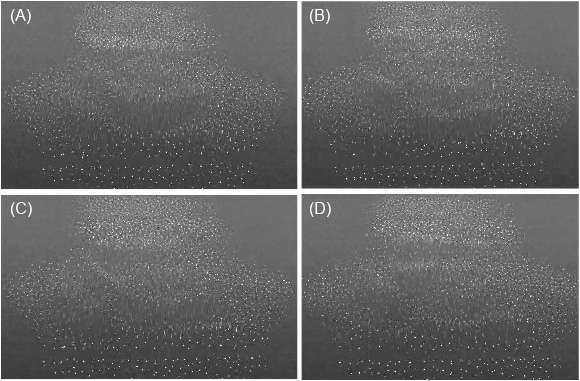}
\caption{A movie of dust particles in the MDPX device recorded at 12.5 frames per second.  The experiment is performed using 2 micron diameter silica particles in an argon plasma at a magnetic field of B = 2 T. Motions of individual particle give rise to collective dust acoustic mode.} 
\label{fig:Auburn1}
\end{figure}

We include a movie example of a dusty plasma created in the Magnetized Dusty Plasma Experiment (MDPX) device at Auburn University, shown in Fig.~\ref{fig:Auburn1}. MPDX  is a superconducting, multi-configuration, high magnetic field (B$_{max} \sim$ 4 Tesla) research instrument.  It is assembled from two main components:  a large cryostat which contains the magnetic field coils and a removable vacuum chamber.  The design and construction of the MDPX device has been discussed extensively in a number of earlier works~\cite{TBL:2014,TKA:2015}.  A key feature of the MDPX design is that the cryostat has central, cylindrical warm bore that has overall dimensions of:  50 cm inner diameter, 122 cm outer diameter, and an overall length of 157 cm.  The cryostat is `split' into upper and lower halves that are 69 cm long with a gap of 19 cm between the two halves.  Each half of the cryostat contains two superconducting coils.  In combination, the four coils can be operated independently so that a variety of magnetic field configurations – from uniform to cusp-like – can be formed.  For the experiments discussed in this paper, the uniform configuration is used.  

\subsection{Atmospheric plasmas}
\begin{figure}[htb]
\includegraphics[width=3.0in, angle=0]{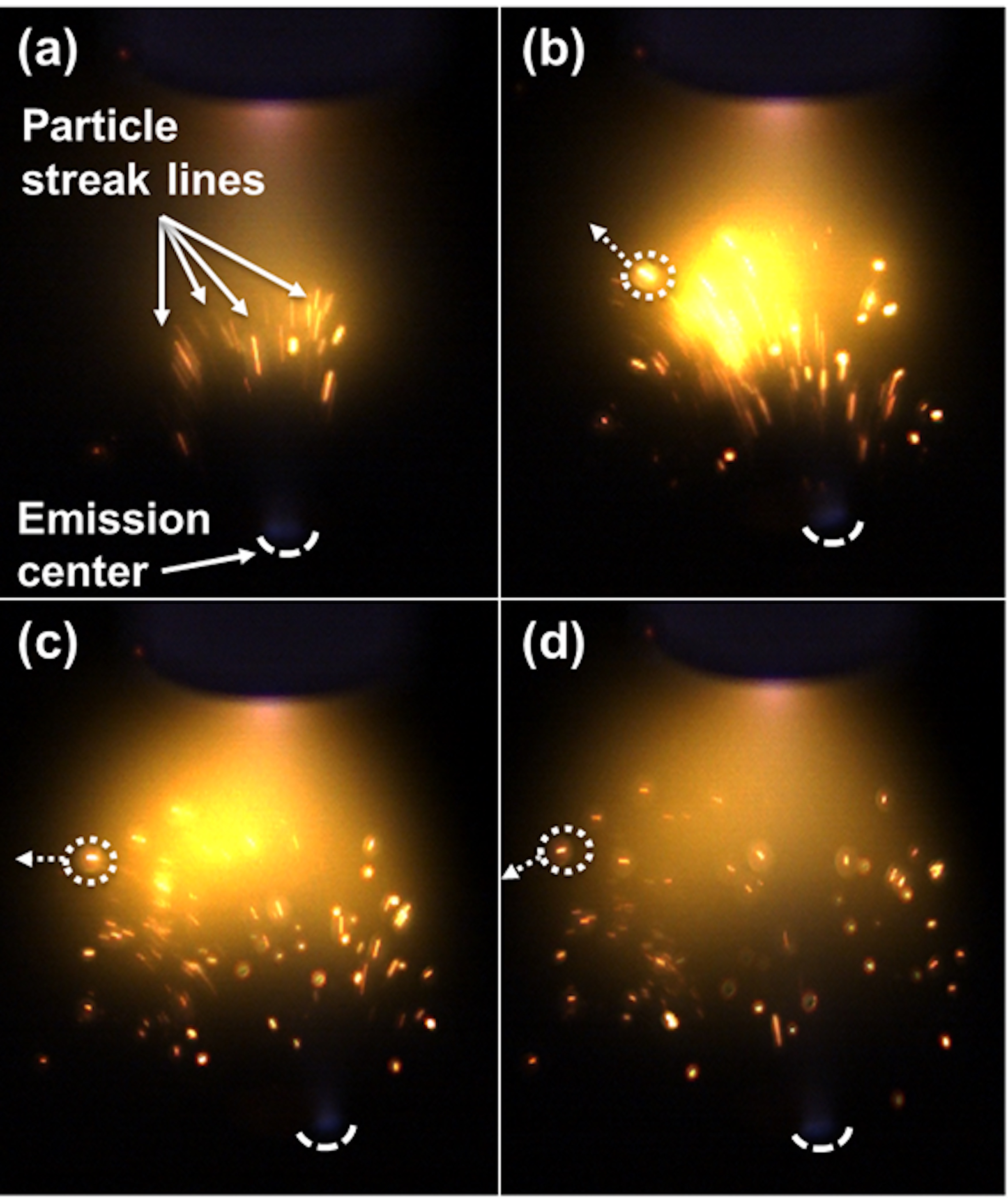}
\caption{Time-resolved particle emission with a high-speed camera at 2k fps: (a-d) t = 2, 4, 6, 8 ms respectively.}
\label{fig:UMich3}
\end{figure}

Atmospheric pressure plasmas (APPs) are non-equilibrium plasmas produced at the ambient atmosphere. These plasmas are currently being investigated for a number of applications ranging from wound healing to material processing to water purification~\cite{Mich1, Mich2, Mich3}.  A particular class of these discharges is the 1 ATM DC glow with a liquid anode. Here the anode electrode is an electrolytic solution and ions complete the electric current flow.  In such discharges, the anode attachment can self-organize into complex shapes observable at the liquid surface. These discharges produce copious amounts of nanoparticles in the liquid phase.  Under certain conditions, particles and droplets derived from the liquid phase are injected into the plasma above. Upon injection, metal nanoparticles are rapidly oxidized, appearing as luminous streaks in photographs, as shown in Fig.~\ref{fig:UMich3}(a).   

Dynamic motion of particle swarms can also been seen in Fig.~\ref{fig:UMich3} when the anode attaches to the plasma region above.   A high-speed camera at 2k fps was used to video-record the emission process. As can be seen from the video, the particle emission coincides with the plasma attachment center at the liquid surface. Although the discharge is steady DC on average, the emission of particles appears as cyclic bursts. The bursts release copious amounts of fast moving particles, which appear as streaks due to the insufficient camera temporal resolution at 2k fps.  Additional excitation is apparent as particles reach the core of the discharge as inferred from the over exposed glow region shown in frames Fig.~\ref{fig:UMich3}(b) and (c).  Following the burst release, the particles travel ballistically as inferred from the observed parabolic trajectories. Note that the streaks are longer during the emission phase (frames a, b) in comparison to later stages where the particles follow what appears to be simple rectilinear motion.  This disparity in streak length suggests that the particles emitted are moving considerably faster upon launch.  For non-viscous ballistic motion however the particles velocity at the surface upon return should equal the launch speed, which suggests that at later times, the streaks should have lengths similar to the initially emitted particles.  The absence of the long streaks at later times implies one or more of the following possible scenarios: 1.) A distribution of particles of differing sizes and velocities are launched; 2.) The smaller faster moving particles may move out of the camera focus and do not come near the emission zone; and 3.) Viscosity is not negligible as in the exploding wire experiment~\cite{CWW:2018,SWR:2018}.

The mechanism for the particle injection into the gas phase and even the excitation processes along the trajectories are still not well understood. By capturing the particle swarms with even higher speed cameras as in the exploding wire experiment, and through stereoscopic imaging from different vantage points, we plan to examine the kinematics of the particle generation and motion in further details, including applications of the physics, statistical and data models described here.  

\section{\label{sec:PCIA} Analyses and discussions}
\begin{figure}[htb]
\includegraphics[width=3.0in, angle=0]{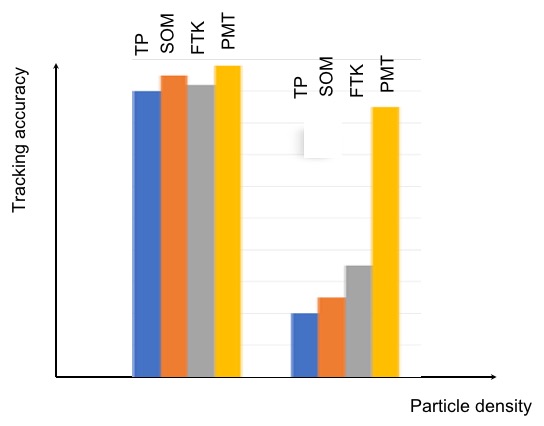}
\caption{Comparison of four different tracking algorithms and their relative accuracy using the exploding wire datasets and varied particle densities.}
\label{fig:algoCC}
\end{figure}

Below we first compare the four algorithms for particle track generation: Trackpy (TP)~\cite{tp:2019}, self-organizng map (SOM), FTK, and physics-constrained motion tracking (PMT), as summarized in Fig.~\ref{fig:algoCC}. Several exploding wire datasets were used at different particle densities. Two datasets are included in Fig.~\ref{fig:EW1}. All four algorithms worked well for low particle densities, achieving an tracking accuracy above 80\%. At higher particle densities, however, only PMT achieved the tracking accuracy above 75\%. In implementation of each of the algorithms, the image analysis workflow can be divided into following modules: image preprocessing (denoising), particle localization and characterization, and particle linking. The overall size of the datasets tends to shrink along the workflow.  The tracking accuracy is defined as the percentage of the particles being assigned to the different tracks with a track length longer than 30. The number of the expected tracks can be estimated with above 95\% accuracy by tallying the number of particles in the video frame that contains the largest number of particles.

\subsection{U-Net for particle localization in noisy videos}
Besides the particle density, another important factor that affects the tracking accuracy is the signal-to-noise ratio (SNR) of the video. The dusty plasma image set has a SNR less than 3, defined as the average particle intensity to the mean background pixel intensity, which is much smaller than that of the exploding wire set (SNR $>$ 6).  We adopted U-Net for particle localization~\cite{Ron:2015,Hao:2019}, a deep CNN (23 convolutional layer total) that can be trained end-to-end with very few images through data augmentation. Here particle localization is equivalent to pixel classification. The U-Net architecture extends the  fully convolutional network~\cite{LSD:2014} and has a symmetric `U' shape. It consists of a contracting first half and an expanding second half~\cite{Ron:2015}. The contracting half has the typical CNN architecture to  capture context.  The symmetric expanding half enables precise localization. 

\begin{figure}[htb]
\includegraphics[width=3.0in, angle=0]{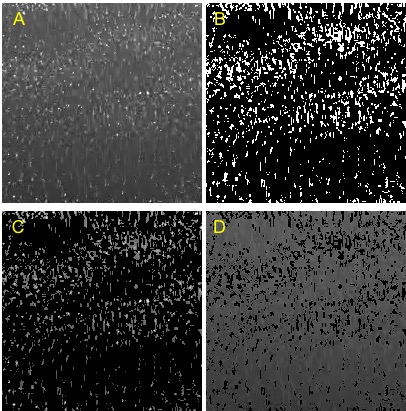}
\caption{ Supervised noise reduction of the dusty plasma image set using U-Net. (A). An original image; (B). U-Net generated binary mask; (C). Masked original image; (D). Background after the subtraction of the masked image (C) from the original image (A).}
\label{fig:Wolfe1}
\end{figure}

The U-Net algorithm here was trained using eighty images from the dusty plasma dataset. The training inputs were produced by extracting 256$\times$256 patches from the frames and upsampling the patches to 512$\times$512. Ground truth segmentations were generated by thresholding the images based upon statistical significance above background pixels. In total this procedure produced 1600 pairs of patches and segmentation maps.  The parameters of U-Net were then optimized by using Adam~\cite{KB:2014} which is an extension of stochastic gradient descent that has an adaptive learning rate. Fig.~\ref{fig:Wolfe1} shows an example processed by the trained U-Net where the segmentation is used to mask the input image.

\begin{figure}[htb]
\includegraphics[width=3.0in, angle=0]{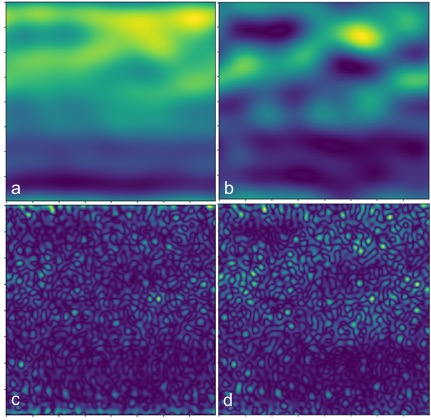}
\caption{Enhancement of SNR using U-Net for different frequency bands: Inverse FFT of the lowest frequencies of the original image (a), the lowest frequencies of U-Net masked image (b), intermediate frequency band of the original image (c), and the intermediate frequency band of the U-Net masked image (d).}
\label{fig:InFourierUNET}
\end{figure}

Another way to look at the workflow of U-Net algorithm is in the Fourier space. Here discrete Fourier transform converts an image into a spacial frequency representation. This representation is complex valued and the modulus (amplitude) can be interpreted as the intensity of a particular frequency in the image. The transform is done efficiently using a fast Fourier Transform (FFT) algorithm~\cite{CT:1965,Swa:1982}.  
U-Net-generated masks allow improved SNR for the dust acoustic mode (the lowest frequency band in the dusty plasma images), Fig.~\ref{fig:InFourierUNET}a \& b, as well as the particle localization (intermediate frequency band), Fig.~\ref{fig:InFourierUNET}c \& d. Wavelet analysis of U-Net outputs gives similar confirmation on SNR improvement.

\subsection{Multiple camera linking}

\begin{figure}[htb]
\includegraphics[width=3.5in, angle=0]{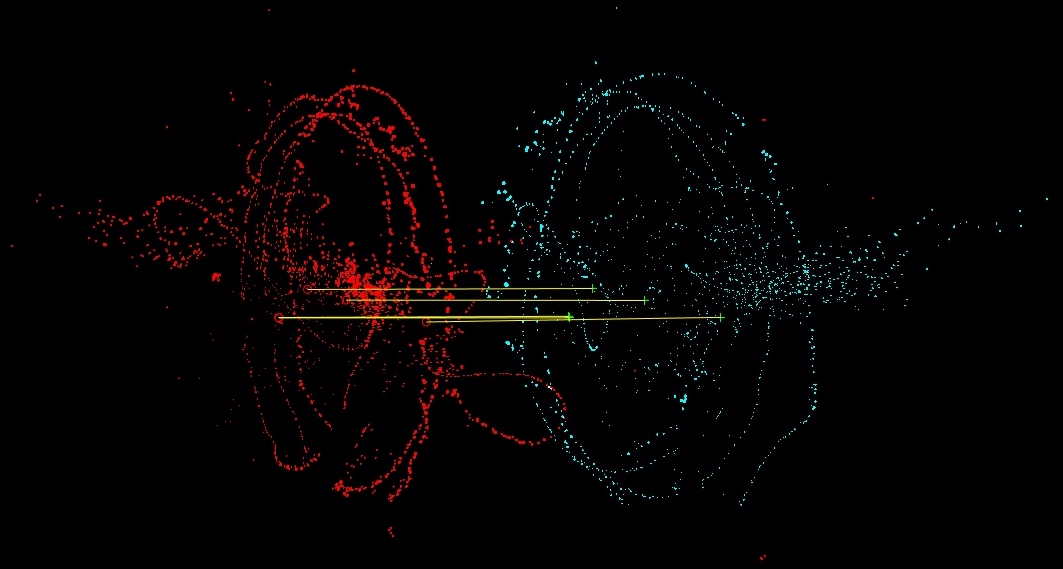}
\caption{An example of scene linking from two camera views. The left and right camera views are shown side by side. The matched points from the two views are linked with lines as shown.}
\label{fig:link1}
\end{figure}

Two or more cameras are used to measure 3D particle positions and motion. Algorithms that link particle positions from different camera views are therefore needed. Groth algorithm~\cite{Gro:1986} is a pattern-matching algorithm for two-dimensional coordinate lists for each camera. The algorithm matches coordinate pairs from two cameras based on the triangles that can be formed from triplets of points in each list. Any translation, rotation, scale change, or flip is not going to change the basic shape of a triangle, although it will change the size and orientation~\cite{Ste:web}.  Each triangle contains two independent shape parameters. In our previous work~\cite{Wang:2016}, we also found that the fast transient events such as particle explosion can be used to match particles in two cameras as well. Examples of pattern matching are shown in Fig.~\ref{fig:link1}. Eight-point algorithms or RANSAC are then used to construct the fundamental matrices for 3D scene reconstruction~\cite{HZ:2004,Wang:2016,SWR:2018}.

\subsection{Trajectory classification}
\begin{figure}[htb]
\includegraphics[width=3.5in, angle=0]{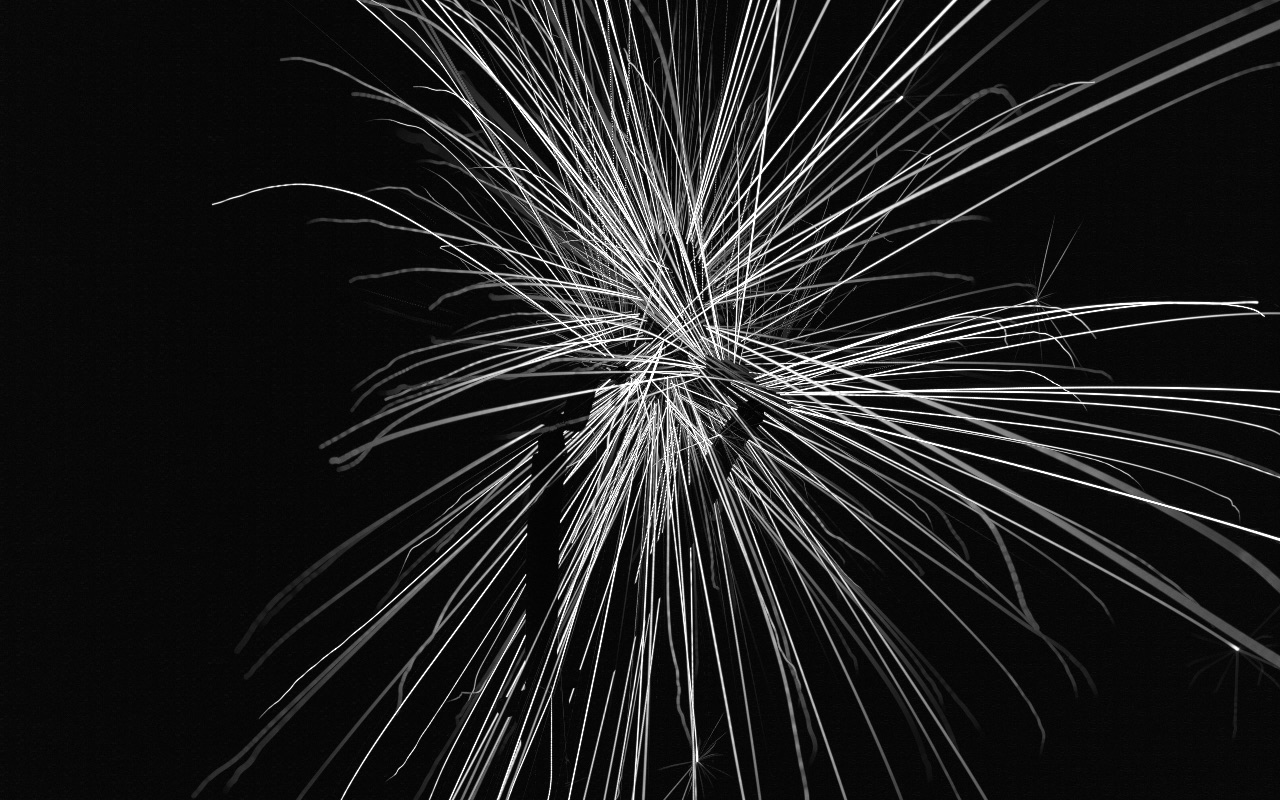}
\caption{ After the background subtraction, possible track candidates from the raw data are obtained from ImageJ/Fiji~\cite{imagej:2019} by summation of the whole video sequence ($\sim$ 1300 frames). }
\label{fig:EWtracks}
\end{figure}

\begin{figure}[htb]
\includegraphics[width=3.5in, angle=0]{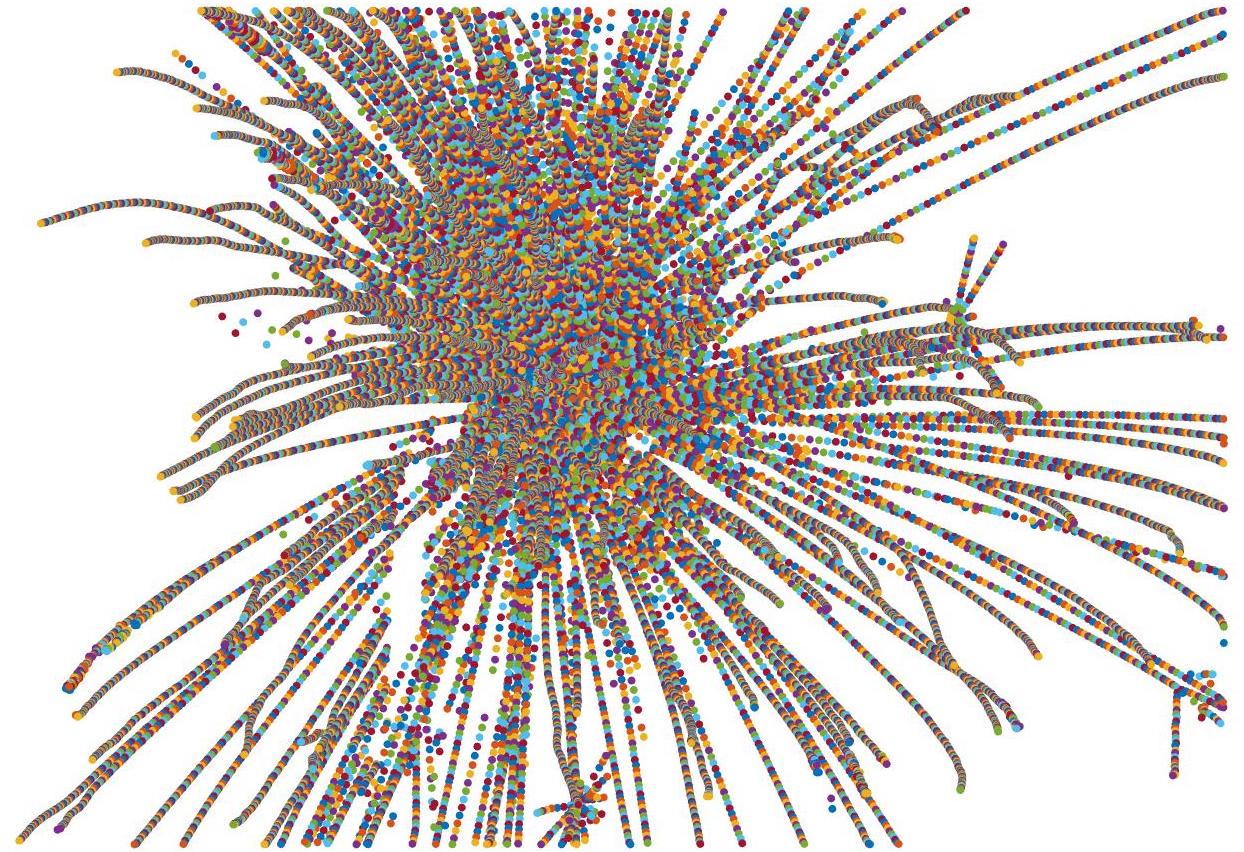}
\caption{Particle tracks recovered from the particle cloud scene using the same dataset as in Fig.~\ref{fig:EWtracks}. The tracks can be characterized as ballistic. Bifurcation of some tracks are due to secondary explosions as discussed previously~\cite{Wang:2016}.}
\label{fig:ftk1}
\end{figure}

Examples of reconstructed tracks are shown in Fig.~\ref{fig:ftk1} for the set of exploding wire data shown in Fig.~\ref{fig:EWtracks}. The combination of gravity and viscous force motivate a second-order polynomial fitting as discussed below. 

\begin{figure}[htb]
\includegraphics[width=3.5in, angle=0]{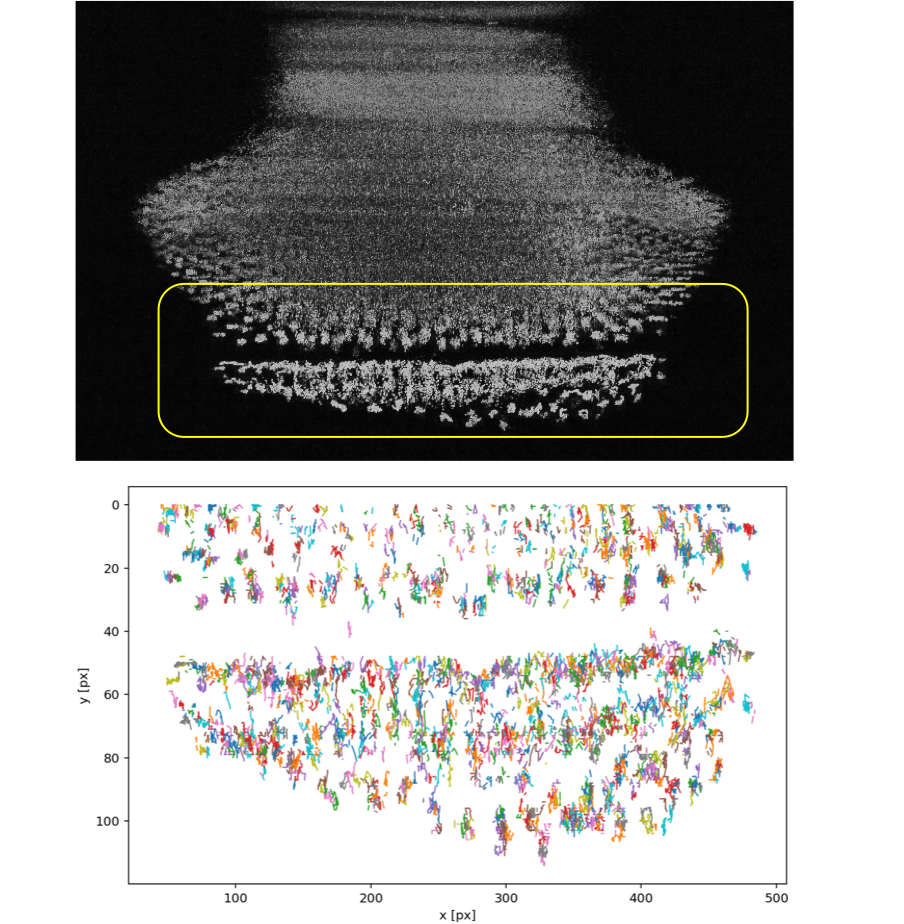}
\caption{ (Top) Summation of 100 video frames from the dusty plasma movie shown in Fig.~\ref{fig:Auburn1} after background subtraction using ImageJ/Fiji. (Bottom) The particle tracks recovered using U-Net from the same region as bounded by the box in the Top. The tracks show a combination of ballistic and diffusive motion.}
\label{fig:DustyT}
\end{figure}

The dusty plasma particle tracking based on U-Net is shown in Fig.~\ref{fig:DustyT}. U-Net provides a labeling as to whether a pixel belongs to a particle or the background. Particle instances are then produced by finding connected components in the segmentation mask. In images, a connected component~\cite{FG:1996} is the largest set of pixels such that they are the same value and there is a path between any two pixels. The components were found using 1-connectivity, meaning that paths do not include diagonals. Each connected component is treated as an individual particle.  The particles are localized by averaging the pixel locations for pixels in each particle. Fig.~\ref{fig:DustyT} shows tracks produced from this localization scheme and a nearest neighbor matching between frames.

The particle tracks from the dusty plasma can be checked against the following model $\Delta l^2 = D \Delta t^k$,
here $\Delta l$ is the distance between the initial time and end time, which differ by $\Delta t$. $k = 1$ corresponds to the normal diffusion with $D$ being the diffusion coefficient. However, most of the particle tracks can not be described by normal diffusion. Some particles are observed to hop its position for a short period of time, similar to ballistic motion, followed by oscillatory motion that can not be reduced to a simple mode. Applications of the U-Net and other algorithms described here and elsewhere to larger datasets will be left for future work. 

\subsection{Visualization of fitting coefficients \label{sec:vfc}}
\begin{figure}[htb]
\includegraphics[width=2.5in, angle=0]{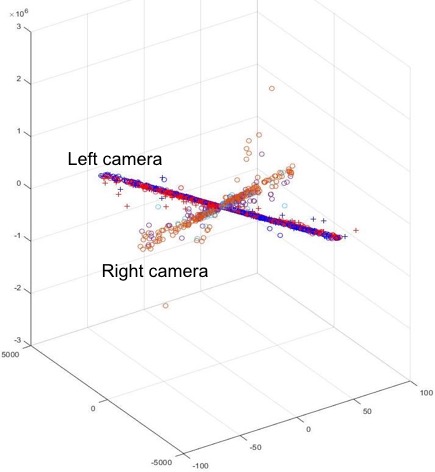}
\caption{\label{fig:DataShape1}  3D scattered plot of the points with the coordinates ($a$, $b$, $c$) as defined by Eq.~(\ref{simp:eq}). The points correspond to the left and right camera are found to cluster differently, as shown in different colors.}
\label{fig:DS}
\end{figure}

Particle tracks allow testing of the different physics models. Previous work on the exploding wire sets motivate a second order polynomial fitting for the trajectories~\cite{Wang:2016,SWR:2018,CWW:2018}, namely $c_1q_x^2+c_2q_xq_y+c_3q_y^2+c_4q_x+c_5q_y+c_6=0$ with fitting coefficients $c_1$ to $c_6$ and camera coordinates $q_x$ and $q_y$. A simplified model 
\begin{equation}
q_y=aq_x^2+bq_x+c
\label{simp:eq}
\end{equation} 
is used here to fit the 2D tracks recovered from FTK as described above in Sec.~\ref{sec:ftk}. Each track has to contain at least 30 data points for their corresponding fitting coefficients. The values of $a$, $b$, and $c$ are assigned to a point as coordinates in the 3D space with three orthogonal axes corresponding to $a$, $b$ or $c$ respectively. The scattered points are shown in Fig.~\ref{fig:DS}. The data points are found to distribute in two main clusters (in two different colors, red and blue). The red cluster corresponds to tracks obtained from the images recorded by the right camera. The blue cluster corresponds to tracks by the left camera.



\section{Summary and perspective}
Particle or object tracking is widely used in astronomy, space science, biology, material science, chemistry, and physics such as fluids and plasmas, and more recently in computer vision, autonomous driving and surveillance. In spite of their contextual and spatial scale differences ranging from galactic sizes to nanometers, similar uses of imaging hardware, similar needs for image processing such as particle/object localization, particle linking from one time to another, particle linking from different camera views, denoising, {\it etc.} have led to open software and algorithms that can be used in multiple fields, making particle tracking a powerful and accessible technique for data-driven plasma science. 

Here we emphasize the microparticle cloud imaging and tracking (mCIT or $\mu$CIT) for plasma science when more than a few micron-size particles can be tracked simultaneously. Rapid recording of individual particle motion leads to fine spatially and temporally resolved data. The whole particle cloud yields spatially extended information as a function of time. The amount of data increases in proportion to the number of particles, the number of cameras, camera sensor size, and camera frame rate. Terabyte of data per day can be readily achievable in today's plasma experiments. Traditional frameworks of data processing based on physics and statistical principles can now be significantly enhanced by data-driven methods, with the potential towards fully automated image processing and information extraction, doing away with {\it ad hoc} assumptions commonly employed in physics and statistical models in order to supplement incomplete information or understanding or to accelerate computation in large scale simulations. Meanwhile, physical and statistical methods are indispensable in the following ways such as providing `ground truths' for data model training, simplifying or constraining data models, which could easily become NP-hard otherwise, and validating data models. 

Three microparticle cloud datasets are presented here: from exploding wires, in dusty plasmas in a strong magnetic field (MDPX) and in atmospheric plasmas. A physics-constrained motion tracker, a Kohonen neural network (KNN) or self-organizing map (SOM), the feature tracking kit (FTK), and U-Net are described and compared with each other for particle tracking using the datasets. Particle density and signal-to-noise ratio have been identified as two important factors that affect the tracking accuracy. Fast Fourier transform (FFT) is used to reveal how U-Net, a deep convolutional neural network (CNN) developed for non-plasma applications, achieves the improvements for noisy scenes. The fitting parameters for  a simple polynomial track model are found to group into clusters that reveal the geometry information about the camera setup. 

Advance and expansion of particle tracking and imaging for plasma science can be expected in conjunction with further maturation of the image processing methods and availability of new data under different plasma conditions. A modular open platform approach to image processing, similar to the popular distribution such as ImageJ/Fiji, will allow community contributions and continuous improvements in time. Such a platform can also combine physics-, statistics-driven algorithms, data methods such as SOM or CNN, novel computing algorithms such as FTK, and additional examples given in the appendix, and lead to hybrid methods that are more accurate or effective than individual elements. Explosive growth of datasets, advances in data methods and algorithms, could also compensate for hardware performances, paving the way towards super-resolution that opens door to nanoparticle and ultimately atom and ion tracking in plasma science.


\begin{acknowledgments}
Los Alamos National Laborotory (LANL) work is supported through Triad National Security, LLC (`Triad') contract \# 89233218CNA000001 by U.S. Department of Energy (DOE), Office of Science, Office of Fusion Energy Sciences (program manager: Dr. Nirmol Podder).  Y.E.K. and J.E.F.  were supported by the U.S. DOE under Grant DE-SC0018058. E.T. is funded by the National Science Foundation (NSF), the NSF EPSCoR program, the US DOE, and the NASA Jet Propulsion Laboratory.  This research used resources of the Magnetized Dusty Plasma Experiment (MDPX), which is a Collaborative Research Facility supported by the Office of Fusion Energy Sciences General Plasma Science program. MPDX was originally funded by the NSF-Major Research Instrumentation program. The feature tracking kit (FTK) development is supported by Laboratory Directed Research and Development (LDRD) funding from Argonne National Laboratory, provided by the Director, Office of Science, of the U.S. Department of Energy under Contract No. DE-AC02-06CH11357. 
\end{acknowledgments}

\appendix
\section{Additional particle tracking resources \label{appen:1}}

We include some additional tracking algorithms and software, intended for beginners and users rather than developers. One open resource is ImageJ/Fiji~\cite{imagej:2019}, which has multiple plug-in particle trackers such as TrackMate, MosaicSuite that includes Particle Tracker tool. OpenPTV is another open source that allows 3D particle velocity field measurement from multiple cameras~\cite{OpenPTV:2019}. Many research groups have developed application-specific trackers. A large fraction of these use MATLAB (licensed) or Python (open) platform, and the codes are downloadable from GitHub or individual research group websites. Searchable examples include TracTrac, Lagrangian Particle Tracking. Reviews and comparative studies exist for cell  and biology applications~\cite{JLM:2008,CSC:2014} and fluid mechanics~\cite{BNK:2019}. The early work on colloidal microscopy by Crocker and Grier led to many development~\cite{CG:1996}. 
For example, Trackpy is a Python implementation of the tracking algorithm originally developed by John Crocker and Eric Weeks in IDL.

\vspace{1cm}

{\bf References}
\bibliography{MCD9}

\end{document}